\documentclass[aip,amsmath,amssymb,reprint]{revtex4-1}
\usepackage{graphicx}
\usepackage{bm}
\usepackage{dcolumn}
\usepackage[utf8]{inputenc}
\usepackage[T1]{fontenc}
\usepackage{mathptmx}

\begin{document}

\title{On the Rayleigh-Taylor unstable dynamics of 3D interfacial coherent \\ structures with time-dependent acceleration}

\author{D.L.~Hill}
\email{des.hill@uwa.edu.au}
\affiliation{University of Western Australia, Perth,WA, 6009, Australia}

\author{A.K.~Bhowmick}
\email{akbhowmi@andrew.cmu.edu}
\affiliation{Carnegie Mellon University, 5000 Forbes Avenue, Pittsburgh, Pennsylvania 15213, USA}

\author{S.I.~Abarzhi}
\email{snezhana.abarzhi@uwa.edu.au}
\affiliation{University of Western Australia, Perth,WA, 6009, Australia}

\date{\today}

\begin{abstract}
Rayleigh-Taylor instability (RTI) occurs in a range of industrial and natural processes. Whereas the vast majority of existing
studies have considered constant acceleration, RTI is in most instances driven by variable acceleration. Here we focus on RTI
driven by acceleration with a power-law time-dependence, and by applying a group theoretic method find solutions to this
classical nonlinear boundary value problem. We deduce that the dynamics is dominated by the acceleration term and that
the solutions depend critically on the time dependence for values of the acceleration exponent greater than $-2$. We find
that in the early-time dynamics, the RTI growth-rate depends on the acceleration parameters and initial conditions. For the
later-time dynamics, we link the interface dynamics with an interfacial shear function, and find a continuous family of regular
asymptotic solutions and invariant properties of nonlinear RTI. The essentially interfacial and multi-scale character of the
dynamics is also demonstrated. The velocity field is potential in the bulk, and vortical structures appear at the interface due
to interfacial shear. The multi-scale character becomes clear from the invariance properties of the dynamics. We also achieve
excellent agreement with existing observations and elaborate new benchmarks for future experimental work.
\end{abstract}

\maketitle

\section{Introduction}

Rayleigh-Taylor instability (RTI) occurs whenever fluids of different densities are accelerated against their density gradient,
and leads to intense interfacial Rayleigh-Taylor (RT) movement of the fluids \cite{rayleigh,davies,landau}. The phenomenon
plays a key roles in many processes, for instance, supernova explosions and inertial confinement fusion, and has received
significant attention over the past seven decades. A reliable theory of RTI is required in order to expand our knowledge of
non-equilibrium dynamics, to better understand RT-relevant phenomena, and to aid the development and improvement of
processes in the areas of energy production and the environment, among many others \cite{turbulentmixing,anisimov}.
Here we study the long-standing problem of RTI subject to a variable acceleration and use discrete group theory
to solve the boundary value problem for early- and late-time RT evolution \cite{turbulentmixing,abarzhireview,dynamicsreview}.
We directly link the interface dynamics and interfacial shear, discern its invariance properties for a broad range of
acceleration parameters, and reveal the interfacial and multi-scale character of RT dynamics. We also elaborate extensive
theory benchmarks that can be used in future experiments and simulations.

RT flows, whilst occurring in distinct physical circumstances, have similar evolution features
\cite{davies,landau,anisimov,abarzhireview,dynamicsreview,meshkov,meshkov2013,robey}. RTI develops when
the flow fields and/or the interface are slightly perturbed from their equilibrium state \cite{rayleigh}. The interface is transformed
into a composition of small-scale shear driven vortical structures and a large-scale coherent structure of bubbles and spikes,
with a bubble (spike) being a portion of the light (heavy) fluid penetrating into the heavy (light) fluid
\cite{davies,landau,anisimov,abarzhireview,dynamicsreview,meshkov,meshkov2013,robey,kadau,glimm,youngs}. Intense
interfacial fluid mixing ensues with time \cite{anisimov,abarzhireview,meshkov2013,robey,kadau,glimm,youngs}.

RTI and RT mixing are challenging to study in experiments and simulations, and to investigate theoretically
\cite{turbulentmixing,anisimov}. RT experiments in fluids and plasmas use advanced technologies to meet tight requirements
for flow implementation, diagnostics and control \cite{meshkov,meshkov2013,robey}. RT simulations employ highly accurate
numerical methods and massive computations to track unstable interfaces, capture small-scale processes and permit a large
span of scales \cite{kadau,glimm,youngs}.

Advanced theories allow us to better understand non-equilibrium RT dynamics, identify universal properties of asymptotic
solutions, and capture symmetries of RT flows
\cite{chandrasekhar,kull,gauthier,nishihara,garabedian,inogamov,abarzhisteady,anr}. Significant success has been recently
achieved in the understanding of RTI and RT mixing with constant acceleration \cite{turbulentmixing,anisimov}. In particular,
the group theory approach has elaborated the multi-scale character of nonlinear RTI, thereby explaining earlier observations
\cite{anisimov,abarzhireview,dynamicsreview,meshkov,meshkov2013,robey,kadau,glimm,youngs}.

Here we study RTI subject to a variable acceleration. Only limited information is currently available on RT dynamics under
these conditions, suggesting the need for a systematic approach \cite{turbulentmixing,supernovae}. We consider
accelerations with power-law time-dependence. These are important to study because they may result in new invariant and
scaling properties of the dynamics \cite{sedov}. They can be tuned to better model realistic environments and thus ensure
practicality of our results \cite{turbulentmixing,arnett,haan,peters,rana,buehler,supernovae}.

We consider RTI in a 3D spatially extended periodic flow and apply group theory to solve the relevant boundary value
problem, which involves boundary conditions at the interface and at the outside boundaries
\cite{dynamicsreview,abarzhisteady,anr,supernovae}. For early-time dynamics we identify the dependence of the RTI growth-rate
on the acceleration parameters and initial conditions. For late-time dynamics, we directly link the interface dynamics to
interfacial shear, find a continuous family of regular asymptotic solutions, and discover invariance properties of nonlinear RTI.
The parameters of the critical, Atwood, Taylor and flat bubbles are identified, including their velocities, curvatures, Fourier
amplitudes, and interfacial shear functions. We also reveal the essentially interfacial and multi-scale character of RT dynamics.
The former is exhibited by the velocity field having intense fluid motion near the interface and effectively no motion in the bulk.
The latter follows from the invariance properties of the dynamics set by the interplay of the two macroscopic
length-scales - the wavelength and the amplitude of the interface.

Our theory resolves the long-standing mathematical problem
\cite{chandrasekhar,kull,gauthier,nishihara,garabedian,inogamov,abarzhisteady,anr}, achieves excellent agreement with
available observations \cite{meshkov,meshkov2013,robey,kadau,supernovae,swisher}, and elaborates new benchmarks for
future experiments and simulations, in order to better understand RT-relevant processes
\cite{arnett,haan,peters,rana,buehler,supernovae,swisher}.

\section{The method of solution}

\subsection{The governing equations}

The dynamics of ideal fluids is governed by conservation of mass, momentum and energy:
\[\frac{\partial \rho}{\partial t}+\sum_{i=1}^3\frac{\partial}{\partial x_i}(\rho v_i)=0,\]
\[\frac{\partial}{\partial t}(\rho v_j)+\sum_{i=1}^3\frac{\partial}{\partial x_i}(\rho v_iv_j)+\frac{\partial P}{\partial x_j}=0,\]
\begin{equation} \label{eq:laws}
\frac{\partial E}{\partial t}+\sum_{i=1}^3\frac{\partial}{\partial x_i}((E+P)v_i)=0,
\end{equation}
where $(x_1,x_2,x_3)=(x,y,z)$ are the spatial coordinates, $t$ is time, ($\rho,{\bf v},P,E)$ are the
fields of density $\rho$ , velocity ${\bf v}$ , pressure $P$ and energy $E=\rho(e+\frac{1}{2}{\bf v}^2)$, where
$e$ is the specific internal energy \cite{abarzhisteady}.

We consider immiscible, inviscid fluids of differing densities, separated by a sharp interface. It is required that  momentum must
be conserved at the interface and that there can be no mass flow across it. Hence the boundary conditions at the interface are
\begin{equation} \label{eq:ics}
\left[{\bf v}\cdot{\bf n}\right]=0, \quad [P]=0, \quad
\left[{\bf v}\cdot{\bm \tau}\right]={\rm arbitrary}, \quad \left[w\right]={\rm arbitrary},
\end{equation}
where $[\cdots]$ denotes the jump of functions across the interface; ${\bf n}$ and ${\bm \tau}$ are the normal and
tangential unit vectors of the interface with ${\bf n}=\frac{{\bm \nabla}\theta}{\vert{\bm \nabla}\theta\vert}$ and
${\bf n}\cdot{\bm \tau}=0$; $w=e+\frac{P}{\rho}$ is the specific enthalpy; $\theta=\theta(x,y,z,t)$, is a local scalar
function, with $\theta=0$ at the interface and $\theta>0$ $(\theta<0)$ in the bulk of the heavy (light) fluid, indicated
hereafter by subscript $h(l)$.

The heavier fluid sits above the lighter fluid and the entire system is subject to a time-dependent downwards
acceleration field, directed from the heavy to the light fluid and is the power-law function of time ${\bf g}=(0,0,-g)$ where
$g=Gt^a$. Here $a$ is the acceleration exponent, and $G>0$ is the acceleration pre-factor \cite{nishihara,garabedian,inogamov}.
Their dimensions are $[G]=ms^{-(a+2)}$ and $[a]=1$. This modifies the pressure field: $P\to P+\rho g z$.
We assume that there are no mass sources and hence the boundary conditions
\begin{equation} \label{eq:bcs}
\lim_{z\to\infty}{\bf v}_h={\bf 0}, \hspace{2cm} \lim_{z\to-\infty}{\bf v}_l={\bf 0}.
\end{equation}

There are two natural time scales in the problem, these are $\tau_g=(kG)^{-1/(a+2)}$ and $\tau_0=\frac{1}{kv_0}$, where $v_0$
is some initial growth rate and $1/k$ is the length scale. We consider here acceleration-driven RT dynamics $a>-2$.
In this case, the former time scale is fastest, $\tau_g\ll\tau_0$. We set the time scale to be $\tau=\tau_g$. Time is $t\gg t_0$
with $t_0\gg\tau$, and the Atwood number is $A=(\rho_h-\rho_l)/(\rho_h+\rho_l)$ and $0<A<1$.

\subsection{Large-scale coherent structures}

These are arrays of bubbles and spikes periodic in the plane normal to the acceleration
direction. At such scales the flow can be assumed to be irrotational at these large scales. We also assume that the
fluids are incompressible and hence that the velocities are expressible in terms of scalar potentials $\Phi_h(x,y,z,t)$
and $\Phi_l(x,y,z,t)$. Because the fluids are ideal these will be harmonic. That is, $\nabla^2\Phi_h=0\ {\rm in}\ \theta>0$
and $\nabla^2\Phi_l=0\ \theta<0$.

We focus on bubbles propagating in the $z$-direction. For convenience our calculations are performed in the
frame of reference moving with velocity $v(t)$ in the $z$-direction, where $v(t)=\partial z_0/\partial t$
and $z_0$ are the velocity and position of the bubble in laboratory reference frame. The interface shape is
$\theta(x,y,z,t)=z-z^*(x,y,t)=0$, and the interface conditions are then
\[\rho_h\left(\nabla\Phi_h\cdot{\bf n}+\frac{\dot\theta}{\vert\nabla\theta\vert}\right)=0
=\rho_l\left(\nabla\Phi_l\cdot{\bf n}+\frac{\dot\theta}{\vert\nabla\theta\vert}\right)\]
\[\rho_h\left(\frac{\partial \Phi _h}{\partial t}+\frac{\vert\nabla\Phi_h\vert^2}{2}
+\left(g(t)+\frac{dv}{dt}\right)z\right)\]
\begin{equation} \label{eq:interfaceconds}
=\rho_l\left(\frac{\partial \Phi _l}{\partial t}+\frac{\vert\nabla\Phi_l\vert^2}{2}
+\left(g(t)+\frac{dv}{dt}\right)z\right)
\end{equation}
The vertical far-field boundary conditions are
\begin{equation} \label{eq:farfield}
\frac{\partial \Phi _h}{\partial z}\Big\vert_{z\to\infty}=-v(t), \hspace{1cm}
\frac{\partial \Phi _l}{\partial z}\Big\vert_{z\to-\infty}=-v(t).
\end{equation}

\subsection{The dynamical system}

The periodic nature of the large-scale coherent structure can be accommodated by appealing to the theory of discrete
groups \cite{anisimov,abarzhireview,dynamicsreview}. We first identify groups enabling structurally stable dynamics.
These are, for example, the group $p6mm$ for hexagonal symmetry, $p4mm$ for square symmetry, $p2mm$ for rectangular
symmetry. The relevant symmetry group (in our case, $p4mm$) dictates a specific Fourier series (an irreducible
representation of the group) which can be used to solve the nonlinear boundary value problem Eqs.
\ref{eq:interfaceconds},\ref{eq:farfield}. We then make spatial expansions in the vicinity of the tip of a bubble. This approach
reduces the governing equations to a dynamical system of ordinary
differential equations in terms of interface variables and Fourier moments 
\cite{anisimov,abarzhireview,dynamicsreview,inogamov,abarzhisteady,anr,supernovae}.

For three-dimensional flow with square symmetry, the potentials are
\[\Phi_h(x,y,z,t)=\sum_{m,n=0}^\infty\Phi_{mn}(t)\left(\frac{\cos(mkx)\cos(nky)
e^{-\alpha_{mn}kz}}{\alpha_{mn}k}+z\right),\]
\begin{equation} \label{eq:phiforms}
\Phi_l(x,y,z,t)=\sum_{m,n=0}^\infty\tilde\Phi_{mn}(t)\left(\frac{\cos(mkx)\cos(nky)
e^{\alpha_{mn}kz}}{\alpha_{mn}k}-z\right),
\end{equation}
where $\alpha_{mn}=\sqrt{m^2+n^2}$, $m$ and $n$ are integers, $k=\frac{2\pi}{\lambda}$ is the wavenumber, 
$\Phi_{mn}$ and $\tilde\Phi_{mn}$ are the Fourier amplitudes for the heavy and light fluids respectively, and
$\Phi_{00}=\tilde\Phi_{00}=0$. Symmetry requires that $\Phi_{mn}=\Phi_{nm}$ and $\tilde\Phi_{mn}=\tilde\Phi_{nm}$.

In order to examine the local behavior of the interfacial dynamics in the vicinity of the bubble tip, we expand the
interface function in a power series about $(x,y)=(0,0)$. In the moving frame of reference, this is
\begin{equation} \label{eq:zetaform}
z^*(x,y,t)=\sum_{N=1}^\infty\sum_{i+j=N}\zeta_{ij}(t)x^{2i}y^{2j},
\end{equation}
where $\zeta_{ij}(t)=\zeta_{ji}(t)$ due to symmetry, $\zeta(t)=\zeta_{10}(t)$ is the the principal curvature at the
bubble tip, and $N=i+j$ is the order of the approximation. To lowest order (that is, $N=1$), the interface is
$z^*(x,y,t)=\zeta_1(t)(x^2+y^2)$.

The Fourier series and interface function are substituted into the interface conditions and the resulting expressions
expanded as Taylor series. This yields a system of ordinary differential equations for $\Phi_m(t)$, $\tilde\Phi_m(t)$
and $\zeta_{ij}(t)$. We may express the potentials in terms of moments
$M_{a,b,c}(t)=\sum_{mn}\Phi_{mn}(t)(mk_x)^a(nk_y)^b\alpha_{mn}^c$
and their tilde equivalents. We note that by symmetry, $M_{a,b,c}=M_{b,a,c}$ and
$M_{a+2,b,c}+M_{a,b+2,c}=M_{a,b,c+2}$ and similarly for $\tilde M$.
The vertical far-field conditions give $M_0=-\tilde M_0=-v(t)$. For $N=1$, we abbreviate the series to second order
in $x$ and $y$, and first order in $z$ since $z^*(x,y,t)$ is quadratic in $x$ and $y$. The interface conditions become
\begin{equation} \label{eq:kinematic}
\dot\zeta_1=4M_1\zeta_1+\frac{M_2}{2}, \hspace{2cm}
\dot\zeta_1=4\tilde M_1\zeta_1-\frac{\tilde M_2}{2},
\end{equation}
\[(1+A)\left(\frac{\dot M_1}{2}+\zeta_1\dot M_0-\frac{M_1^2}{2}-\zeta_1g\right)\]
\begin{equation} \label{eq:momentum}
=(1-A)\left(\frac{\dot {\tilde{M_1}}}{2}-\zeta_1\dot {\tilde M_0}-\frac{\tilde M_1^2}{2}-\zeta_1g\right),
\end{equation}
where $M_0=M_{0,0,0}$, $M_1=M_{2,0,-1}$ and $M_2=M_{2,0,0}$. This representation in terms of moments
$M$ and $\tilde M$, and the interface variable $\zeta$, accommodates the nonlocal nature of the nonlinear
dynamics and enables us to investigate the interplay of harmonics and derive regular asymptotic solutions.

\subsection{Asymptotic solutions}

\subsubsection{Early time, $t-t_0\ll\tau$}

In this regime, the system can be linearised and only one harmonic is needed, that is, the moments retain only one
Fourier amplitude.  The initial conditions at time $t_0$ are the initial curvature $\zeta_0=\zeta_1(t_0)$ and velocity
$v_0=v(t_0)$.

For a broad class of initial conditions, integration of the governing equations is a challenge. The solution can be found
\cite{davies,landau,supernovae} when the amplitude of the initial perturbation is small $\tau k\vert v_0\vert\ll1$, and the
interface is nearly flat $\vert\zeta_1/k\vert\ll1$. The system reduces to
\begin{equation} \label{eq:earlytime}
\dot\zeta_1=-\left(\frac{k^2}{4}\right)v, \hspace{1cm} \dot v=-\left(\frac{4A}{k}\right)\zeta_1Gt^a.
\end{equation}

\subsubsection{Later time, $t-t_0\gg\tau$}

In the later-time situation, the behaviour is nonlinear and multiple harmonics must be retained. We find asymptotic
solutions for the relevant equations and determine their stability. To leading order in time, regular asymptotic solutions
will have the following time-dependence:
\[\frac{\zeta_1}{k}\sim{\rm const}, \hspace{2cm}
\Phi_{mn},\tilde \Phi_{mn}\sim\left(\frac{1}{k\tau}\right)\left(\frac{t}{\tau}\right)^\frac{a}{2},\]
\begin{equation} \label{eq:asymforms}
M_j,\ \tilde M_j\sim k^j\Phi_{mn},\ k^j\tilde \Phi_{mn}, \hspace{2cm} j=0,1,2.
\end{equation}

We investigate the stabiity of the asymptotic solutions by including the perturbations
\[\delta\zeta_1(t)\sim \xi(t), \hspace{2cm} \delta\Phi_{mn}(t),\ \delta\tilde\Phi_{mn}(t)\sim\dot\xi(t),\]
\[\quad \delta M_{a,b,c}(t),\ \delta\tilde M_{a,b,c}(t)\sim\dot\xi(t)\]
where $\xi(t)$ is to be determined. Since there are in total four equations (three conservation equations
and the auxilliary condition $M_0=-\tilde M_0$), we consider perturbations of four of the variables (three Fourier
amplitudes and the bubble curvature) to analyze the solution stability. The results are insensitive to which harmonic
we choose to leave unperturbed.

The interface conditions require that $\dot \xi=\beta t^\frac{a}{2}\xi$, where $\beta$ is to be determined. Perturbations
are stable if ${\rm Re}[\beta]<0$, otherwise unstable. The solution of this differential equation is
\begin{equation} \label{eq:asymptotic}
\xi(t)=\exp\left(\frac{2\beta}{a+2}t^\frac{a+2}{2}\right).
\end{equation}
For $a=0$ we have exponential solutions and the family of solutions are exponentially stable/unstable.

For $a=-2$ we have the power-law solution $\xi(t)=t^\beta$ and the family of solutions are asymptotically
(but not exponential) stable/ unstable. Eq. \ref{eq:momentum} now yields a quadratic equation for $\beta$ which
does not depend on the value of $a$, but is algebraically unwieldy. 
\section{Results}

\subsection{The early-time regime}

When $t-t_0\ll\tau=(kG)^{-1/(a+2)}$, only first order harmonics are retained in moments,
that is, $M_0=2\Phi_{10}$, $\tilde M_0=2\tilde\Phi_{10}$; $M_n=k^n\Phi_{10}$, $\tilde M_n=k^n\tilde\Phi_{10}$.
For an almost flat interface the solution is
\[-\frac{\zeta}{k}=C_1\sqrt{\frac{t}{\tau}}I_\nu\left(\frac{\sqrt{A}}{s}~\left(\frac{t}{\tau}\right)^s\right)
+C_2\sqrt{\frac{t}{\tau}}I_{-\nu}\left(\frac{\sqrt{A}}{s}\left(\frac{t}{\tau}\right)^s\right),\]
\begin{equation} \label{eq:almostflat}
v=\frac{4}{k}\frac{d}{dt}\left(-\frac{\zeta}{k}\right).
\end{equation}
where $s=\frac{a+2}{2}$, $\nu=\frac{1}{a+2}$, $I_\nu$ is the modified Bessel function of order $\nu$, and $C_1$ and $C_2$
are integration constants defined by the initial conditions $\zeta_0=\zeta(t_0)$ and $v_0=v(t_0)$ with
$\zeta_0/k\ll1$ and $\tau k\vert v_0\vert\ll1$ \cite{davies,landau,supernovae}. An analysis of the very-early-time
($t\sim\ t_0$) dynamics yields
\[\zeta-\zeta_0\sim-\frac{k^2v_0}{4}(t-t_0),\]
\begin{equation} \label{eq:veryearlytime}
v-v_0\sim-\frac{Akv_0^2}{2}(t-t_0)-4A\left(\frac{\zeta_0}{k}\right)\left(\frac{1}{\tau k}\right)
\left(\frac{t_0}{\tau}\right)^a\left[\frac{t-t_0}{\tau}\right]
\end{equation}
which suggests the positions of bubbles ($\zeta\le0,v\ge0$) and spikes ($\zeta\ge0,v\le0$) are defined by the initial
morphology of the interface, with bubbles formed for $\frac{\zeta_0}{k}<0$ and spikes formed for $\frac{\zeta_0}{k}>0$.

\subsection{The later-time regime}

\begin{figure}
\includegraphics[width=0.75\linewidth]{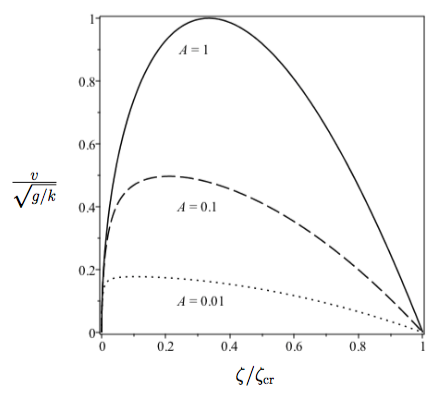}
\caption{Bubble tip velocity as a function of curvature for various Atwood numbers}
\label{fig:tipvelocity}
\end{figure}

\begin{figure}
\includegraphics[width=0.8\linewidth]{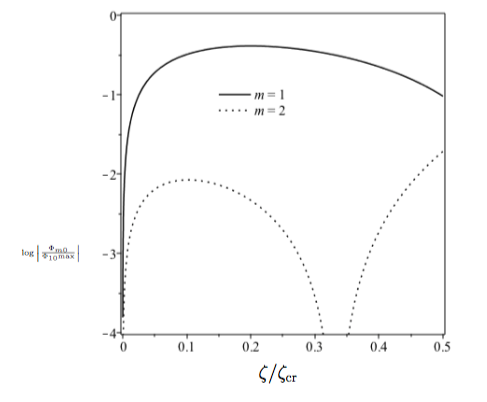}
\caption{Logarithm of absolute values of the $1^{\rm st}$ (solid line) and $2^{\rm nd}$ (dashed line) Fourier amplitiudes in
the heavy fluid}
\label{fig:heavy}
\end{figure}

\begin{figure}
\includegraphics[width=0.8\linewidth]{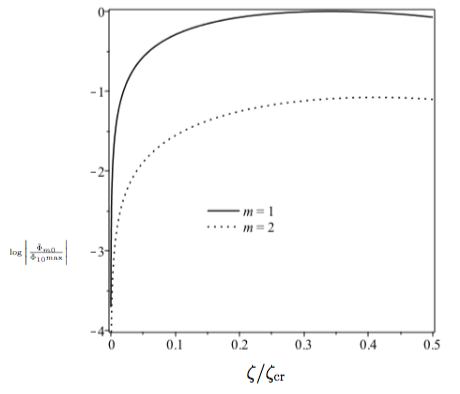}
\caption{Logarithm of absolute values of the $1^{\rm st}$ (solid line) and $2^{\rm nd}$ (dashed line) Fourier amplitiudes in
the light fluid}
\label{fig:light}
\end{figure}

At later times, spikes are singular (the singularity is finite-time), whereas bubbles are regular
\cite{anisimov,abarzhireview,dynamicsreview}. For $t\gg\tau$, higher order harmonics are retained in the expressions for the
moments, and regular asymptotic solutions can be derived. For $N=1$, the first two harmonics are retained and we arrive
at a one-parameter family of solutions (as there are four equation in five unknowns). We choose the bubble curvature
$\zeta$ to parametrize the family.  Substitution of the asymptotic forms Eq. \ref{eq:asymforms} into Eqs. \ref{eq:kinematic} and
\ref{eq:momentum}, employing a dominant balance agrument and solving the resulting set of equations leads to the solution
\begin{equation} \label{eq:nonlinear}
v(t)=\frac{9-64\sigma^2}{k\tau}\sqrt{\frac{2A\sigma}{64A\sigma^2+9A+48\sigma}}
\left(\frac{t}{\tau}\right)^\frac{a}{2},\quad \sigma=-\frac{\zeta}{k}.
\end{equation}
which is valid for $\sigma\in(0,\sigma_{\rm cr})$ where $\sigma_{\rm cr}=\frac{3}{8}$ with corresponding
$\zeta_{\rm cr}=-\frac{3}{8}k$. Fig. \ref{fig:tipvelocity} shows the bubble tip velocity as a function of the bubble curvature.
We observe that the bubble tip velocity is larger for larger values of the Atwood number $A$, and occurs at a steeper
curvature. The Fourier amplitudes are
\[\Phi_{10}=\frac{-8\sigma+2}{8\sigma-3}v, \quad \Phi_{20}=\frac{8\sigma-1}{16\sigma-6}v,\]
\[\tilde\Phi_{10}=\frac{8\sigma+2}{8\sigma+3}v, \quad \tilde\Phi_{20}=\frac{-8\sigma-1}{16\sigma+6}v,\]
\begin{equation} \label{eq:phisandv}
-v=2\Phi_{10}+2\Phi_{20}, \hspace{1cm} v=2\tilde\Phi_{10}+2\tilde\Phi_{20}.
\end{equation}
Solutions for $N>1$ can likewise be calculated. These solutions converge for increasing $N$ and in each case the
lowest order harmonics are dominant. Figs. \ref{fig:heavy} and \ref{fig:light} demonstrate that the second Fourier
amplitude is much smaller than the first for $\sigma<\sigma_{\rm cr}$.

At $N=1$, the solution with maximum growth rate is a curved bubble, having a curvature which depends on the Atwood number. 
The curvature $\sigma_{\rm max}$ which maximizes the velocity satisfies
\[\sigma_{\rm max}^4+\frac{1}{A}\sigma_{\rm max}^3+\frac{9}{32}\sigma_{\rm max}^2-\left(\frac{3}{16}\right)^3=0\]
and the corresponding maximum velocity is
\begin{equation} \label{eq:vmax}
v_{\rm max}(t)=\left(\frac{1}{k\tau}\right)\left(\frac{t}{\tau}\right)^\frac{a}{2}(8\sigma_{\rm max})^\frac{3}{2}.
\end{equation}
This is a universal relation between the curvature and the maximum velocity. 

\subsection{The effect of shear}

\begin{figure}
\includegraphics[width=0.75\linewidth]{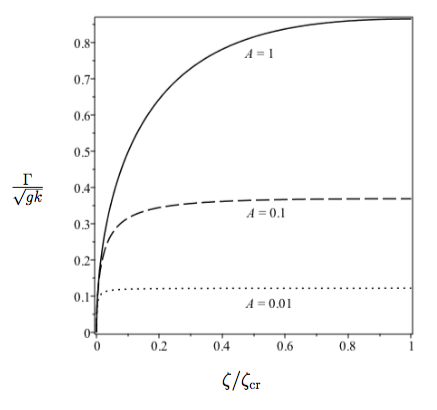}
\caption{Shear as a function of curvature for various Atwood numbers}
\label{fig:shear}
\end{figure}

\begin{figure}
\includegraphics[width=0.75\linewidth]{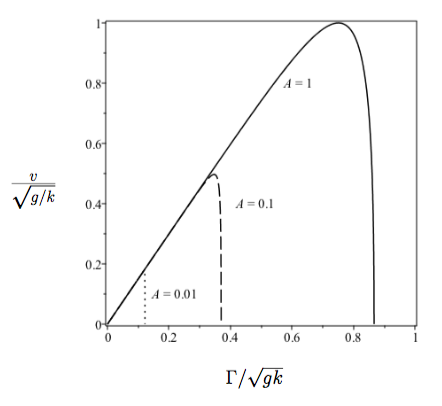}
\caption{Bubble tip velocity as a function of shear for various Atwood numbers}
\label{fig:shearvel}
\end{figure}

The multiplicity of these solutions is also due to the presence of shear at the interface. We define shear function $\Gamma$
to be the spatial derivative of the jump in the tangential velocity across the interface. We find that in the vicinity of
the bubble tip it is $\Gamma=\tilde M_1-M_1$. Specifically,
\begin{equation} \label{eq:interfaceshear}
\Gamma(t)=\frac{6kv(t)}{9-64\sigma^2}
=\frac{6}{\tau}\sqrt{\frac{2A\sigma}{64A\sigma^2+9A+48\sigma}}\left(\frac{t}{\tau}\right)^\frac{a}{2}
\end{equation}
and is a strictly monotone function of $\sigma$, rising from $\Gamma=0$ at $\sigma=0$ to
$\Gamma=\Gamma_{\rm max}$ at $\sigma=\sigma_{\rm cr}$. Fig. \ref{fig:shear} shows the interface shear as a function
of the bubble curvature. We note that the shear function is larger for larger values of the Atwood number $A$, and in each case
tends towards a constant value as the curvature increases. Fig. \ref{fig:shearvel} shows the variation of the bubble tip velocity
with the interface shear function. We note that the velocity achieves a maximum at some point and that this peak occurs at higher
values of the shear function for larger Atwood numbers. The peak velocity itself is also larger for larger Atwood numbers.
We particularly note that very soon after the velocity curve reaches its peak, it drops sharply. This rapid change presents
significant problems for numerical simulations of RTI.

\subsection{Special solutions}

\subsubsection{The flat bubble}

The solution corresponding to a flat bubble is simply $\sigma_f=0$, $v_f=0$, $\Gamma_f=0$, $\Phi_{10{\rm f}}=0$,
$\Phi_{20{\rm f}}=0$, $\tilde\Phi_{10{\rm f}}=0$, $\tilde\Phi_{20{\rm f}}=0$.

\subsubsection{The Atwood bubble}

The fastest member of the family we refer to as the `Atwood bubble' to emphasise its complex dependence on the
Atwood number. The solution is
\begin{equation} \label{eq:atwood}
\sigma_{\rm A}=\sigma_{\rm max}, \hspace{.75cm} v_{\rm A}=v_{\rm max}(t), \hspace{.75cm}
\hat\Gamma_{\rm A}=\frac{6V_{\rm A}}{9-{V_{\rm A}}^\frac{4}{3}},
\end{equation}
where
\[V_{\rm A}=k\tau\left(\frac{\tau}{t}\right)^{-\frac{a}{2}}v_{\rm A}(t), \hspace{1cm}
\hat\Gamma_{\rm A}=\tau\left(\frac{\tau}{t}\right)^{-\frac{a}{2}}\Gamma_{\rm A}(t),\]
and the Fourier amplitudes are given by Eq. \ref{eq:phisandv}. We note that the $\hat\Gamma_{\rm A}(V_{\rm A})$
function is very nearly linear for $V_{\rm A}\in(0,1)$.

In the limit $A\to1$, the solution is
\[\sigma_{\rm A1}=\frac{1}{8}-\frac{1-A}{64}, \hspace{1cm}
v_{\rm A1}=\frac{1}{k\tau}\left(1-\frac{3-3A}{16}\right)\left(\frac{t}{\tau}\right)^\frac{a}{2},\]
\begin{equation} \label{eq:atwood1}
\Gamma_{\rm A1}=\frac{3}{4\tau}\left(1-\frac{7-7A}{32}\right)\left(\frac{t}{\tau}\right)^\frac{a}{2}
\end{equation}
and the Fourier amplitudes $\Phi_{10{\rm A1}}$, $\Phi_{20{\rm A1}}$, $\Phi_{10{\rm A1}}$, $\tilde\Phi_{20{\rm A1}}$
are, respectively,
\[\frac{1}{k\tau}\left(-\frac{1}{2}+\frac{1-A}{16},\frac{1-A}{32},\frac{3}{4}-\frac{19-19A}{128},-\frac{1}{4}+\frac{7-7A}{128}\right)\left(\frac{t}{\tau}\right)^\frac{a}{2}.\]

In the limit $A\to0$, the solution is
\begin{equation} \label{eq:atwood0}
\sigma_{\rm A0}=\frac{3}{16}A^\frac{1}{3}, \hspace{.5cm}
v_{\rm A0}=\frac{1}{k\tau}\sqrt{\frac{27A}{8}}\left(\frac{t}{\tau}\right)^\frac{a}{2}, \hspace{.5cm}
\Gamma_{\rm A0}=\frac{2k}{3}v_{\rm A}.
\end{equation}
and the Fourier amplitudes are $\Phi_{10{\rm A0}}=-\frac{2}{3}v_{\rm A0}$, $\Phi_{20{\rm A0}}=\frac{1}{6}v_{\rm A0}$,
$\tilde\Phi_{10{\rm A0}}=\frac{2}{3}v_{\rm A0}$ and $\tilde\Phi_{20{\rm A0}}=-\frac{1}{6}v_{\rm A0}$.

\subsubsection{The `Taylor' bubble}

We refer to this bubble as a `Taylor bubble' since its curvature is as in \cite{davies} except for a difference
in the wavenumber value. The solution is
\begin{equation} \label{eq:taylor}
\sigma_{\rm T}=\frac{1}{8}, \hspace{1cm} v_{\rm T}=\frac{1}{k\tau}\sqrt{\frac{8A}{5A+3}}\left(\frac{t}{\tau}\right)^\frac{a}{2},
\hspace{1cm} \Gamma_{\rm T}=\frac{3k}{4}v_T.
\end{equation}
The corresponding Fourier amplitudes are $\Phi_{10{\rm T}}=-\frac{1}{2}v_{\rm T}$, $\Phi_{20{\rm T}}=0$,
$\tilde\Phi_{10{\rm T}}=\frac{3}{4}v_{\rm T}$ and $\tilde\Phi_{20{\rm T}}=-\frac{1}{4}v_{\rm T}$. Note that
$\Phi_{20{\rm T}}\ne0$ for $n>1$.

\subsubsection{The critical bubble}

The solution is
\begin{equation} \label{eq:critical}
\sigma_{\rm cr}=\frac{3}{8}, \hspace{.5cm} v_{\rm cr}=0, \hspace{.5cm}
\Gamma_{\rm cr}=\Gamma_{\rm max}=\frac{1}{\tau}\sqrt{\frac{3A}{2(A+1)}}\left(\frac{t}{\tau}\right)^\frac{a}{2}.
\end{equation}
and Fourier amplitudes are $\Phi_{10{\rm cr}}=\frac{\Gamma_{\rm max}}{k}$,
$\Phi_{20{\rm cr}}=-\frac{\Gamma_{\rm max}}{k}$, $\tilde\Phi_{10{\rm cr}}=0$, $\tilde\Phi_{20{\rm cr}}=0$.

\subsection{Stability}

\begin{figure}
\includegraphics[width=0.75\linewidth]{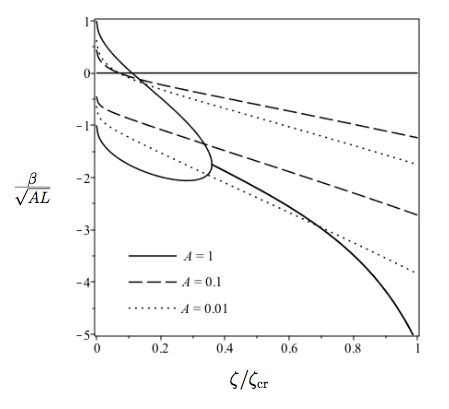}
\caption{Stability profiles for various Atwood numbers and $a>0$}
\label{fig:beta}
\end{figure}

Fig \ref{fig:beta} shows stability profiles for various Atwood numbers and $a>0$. We see that the  flat bubble is unstable,
having $\sigma=\pm\sqrt{AkG}$. In the limits $A\to1^-$ and $A\to0$ stability is achieved when the curvature reaches
\begin{equation} \label{eq:stabilitylimits}
\sigma_{\rm st1}=\frac{1}{24}-\frac{1-A}{250}, \quad
\sigma_{\rm st0}=\frac{-9A+3\sqrt{25A^2+8A}}{64(2A+1)}\sim\frac{3\sqrt{2}}{32}\sqrt{A},
\end{equation}
respectively. All bubbles are stable for $A\ge\frac{1}{24}$ and hence the Taylor and critcal bubbles are stable. The Atwood
bubble is stable when $A=1$,  having $\sigma=-\frac{4}{3}\sqrt{kG},-2\sqrt{kG}$. The Atwood bubble for $A=0.1$ and
$A=0.01$ are also stable and it would appear that the Atwood bubble is stable for any value of Atwood number $A$ at $N=1$.
The $N>1$ analysis is to be presented elsewhere. As is the case for $a=0$, we expect that the stability interval will narrow
sharply for $N>1$.

When $a=0$ and $0\ll A\le1$, velocities $v_{\rm A}$ and $v_{\rm T}$ are close to that of the so-called Layzer-type bubble
$v_L=\frac{1}{k\tau}\sqrt{\frac{2A}{1+A}}$, with which experiments and simulations are often usually compared
\cite{anisimov,abarzhireview,dynamicsreview,inogamov,abarzhisteady,anr,layzer,alon}. Thus, our results excellently agree with
existing observations \cite{meshkov,meshkov2013,robey,kadau,glimm,youngs,supernovae}. Our theory is focused
on large-scale dynamics and presumes that interfacial vortical structures are small-scale. This assumption is applicable
for fluids very different densities and with finite density ratios. For fluids with very similar densities $0\approx A\ll1$ other
approaches can be employed \cite{anisimov,abarzhireview,dynamicsreview,abarzhisteady,anr,supernovae}.

\subsection{The velocity field}

By accurately accounting for the interplay of harmonics and systematically connecting the interfacial velocity and shear
for a broad range of acceleration parameters, we have found that RT dynamics is essentially interfacial: It has intense
fluid motion in the vicinity of the interface, effectively no motion away from the interface and shear-driven vortical structures
at the interface. The velocity is potentisl in the bulk of each fluid. This velocity pattern is observed in experiments and
simulations, demonstrating excellent agreement with our results
\cite{meshkov,meshkov2013,robey,kadau,glimm,youngs,supernovae}.

\section{Discussion}

We have found solutions for acceleration driven RTI in a 3D spatially extended periodic flow in both the early-time regime
and later-timer regime. In the early-time regime the dynamics is faster than exponential for $a>0$ and slower than
exponential for $-2<a<0$. In the later-time regime, bubbles are accelerated for $a>0$, steady at $a=0$, and decelerated
for $-2<a<0$. Since the dependencies are given by standard formulae, we can compare various acceleration exponents.
The dynamics for small exponents are usually a diagnostic challenge becaue it is slow. Specific examples are fusion,
supernovae and nano-fabrication. The dynamics for fast exponents can be easily diagnosed, and the deduced properties
can be applied to the slow-dynamics case. Hence we have obtained an important practical result.

The nonlinear bubble velocity and shear, when rescaled as $k\tau(t/\tau)^{-a/2}v$ and  $\tau(t/\tau)^{-a/2}\Gamma$,
depend only on the interface morphology and flow symmetry. Hence, by analyzing properties of RT bubbles for fast
dynamics and large exponents $a>0$, we can obtain properties of those for slow dynamics and small exponents $-2<a<0$ .
These are especially convenient for studies of RTI in high energy density plasmas in astrophysics and fusion, where RTI is
driven by an explosion or an implosion with blast wave acceleration exponents $-2<a<-1$ in these cases
\cite{sedov,supernovae,swisher}.

Our analysis also elaborates upon diagnostic quantities which have not been discussed before. These are the velocity and
pressure fields, the interface morphology and bubble curvature, the interfacial shear and its link to the bubble velocity and
curvature, the spectral properties of the velocity and pressure, along with the interface growth and growth-rate. By
determining the dependence of these quantities on the density ratio, flow symmetry, and acceleration exponent and strength,
and by identifying their universal properties, and comparing all of these to data obtained for real fluids, we can further
advance our knowledge of RT dynamics in realistic environments, thereby achieving a better understanding of RT relevant
processes and also improve methods of numerical modeling and experimental diagnostics of interfacial dynamics in fluids,
plasmas and other materials.

Our techniques can also be used to study Richtmyer-Meshkov instability (that is, $a<-2$). The results obtained are
very different from those of the Rayleigh-Taylor instability (that is,  $a>-2$).

We note also that the expressions obtained in the constant acceleration $a=0$ case are unexpectedly very similar in form
to those of the $-2<a<0$ case.

\section{References}

\bibliographystyle{apsrev4-1}
\bibliography{bb}

\end{document}